\documentclass[useAMS,usenatbib]{mn2e}

\usepackage{graphicx}

\title[The TP-AGB in hierarchical galaxy formation models]
      {The impact of TP-AGB stars on hierarchical galaxy formation models}

\author[C. Tonini et al.]
{Chiara Tonini$^{1}$
\thanks{E-mail:chiara.tonini@port.ac.uk},
Claudia Maraston$^{1}$, 
Julien Devriendt$^{2}$, 
Daniel Thomas$^{1}$ 
\newauthor
and Joseph Silk$^{2}$ \\
$^{1}$Institue of Cosmology and Gravitation, University of Portsmouth, PO1 3FX Portsmouth, UK\\
$^{2}$University of Oxford, OX1 3PU Oxford, UK\\
}

\begin{document}



\maketitle

\begin{abstract}
The spectro-photometric properties of galaxies in galaxy formation
models are obtained by combining the predicted history of star formation and mass
accretion with the physics of stellar evolution through stellar
population models. In the recent literature, significant differences have emerged regarding
the implementation of the Thermally-Pulsing
Asymptotic Giant Branch phase of stellar evolution.
The emission in the TP-AGB phase 
dominates the bolometric and near-IR spectrum of 
intermediate-age ($\sim 1$ Gyr) stellar populations, hence it is crucial for 
the correct modeling of the galaxy luminosities and colours.
In this paper for the first time, we incorporate a full prescription of 
the TP-AGB phase in a semi-analytic model of galaxy formation.
We find that the inclusion of the TP-AGB in the model spectra 
dramatically alters the predicted colour-magnitude 
relation and its evolution with redshift.
When the TP-AGB phase is active, 
the rest-frame $V-K$ galaxy colours are redder by almost $2$ magnitudes
in the redshift range $z \sim 2-3$ and by $1$ magnitude at $z \sim 1$.
Very red colours are produced in disk galaxies, 
so that the $V-K$ colour distributions of disk and spheroids are virtually
undistinguishable at low redshifts.
We also find that the galaxy K-band emission is
more than $1$ magnitude higher in the range $z \sim 1-3$.
This may alleviate the difficulties met by the hierarchical clustering scenario 
in predicting the red galaxy population at high redshifts. The comparison
between simulations and observations have to be revisited in the light
of our results.

\end{abstract}


\section{Introduction}

The modern theory of the formation and evolution of galaxies   
is built on the CDM hierarchical structure formation paradigm, 
and is supported by two powerful tools, semi-analytic models and
smoothed particle hydrodynamics (SPH) simulations. In both cases, they combine the
results of N-body simulations, producing the assembly of dark matter 
structures, with the baryonic physics describing cooling, star formation,
feedback, chemical enrichment and dynamical interactions. 
The interface between models and observations is mainly 
constituted by the galaxy spectra, which can be modeled and interpreted
to infer mass, age, chemical composition, star formation rate and 
mass assembly history of the galaxy populations (see Baugh, 2006, for a 
recent review). 

The spectra of galaxies are obtained using stellar population 
models, that predict the spectral energy distribution (SED) of ensembles of 
stars as a function of
age, metallicity and the Initial Mass Function (IMF).   
At any given galaxy age, the predicted emitted spectrum is  
a superposition of the synthetic spectra of the Single Stellar Populations 
(SSPs) that compose the total stellar mass, 
assembled according to the model's star formation rate (SFR) as a function of time. 

Stellar population models currently used in published galaxy formation models 
and SPH simulations include Bruzual \& Charlot (2003;
adopted by Bower et al. 2006, Croton et al. 2006, De Lucia et al. 2006, 
Khochfar et al. 2007, Menci et al. 2006, Naab et al. 2007, Nagamine et al. 2005, 
Somerville et al. 2008), GRASIL by Silva et al. (1998; for 
instance Granato et al. 2004, Monaco et al. 2007), STARDUST by 
Devriendt et al. (1999; see Hatton et al. 2003) and
PEGASE by Fioc \& Rocca-Volmerange (1997; 
see Cattaneo et al. 2005, Mori \& Umemura 2006). 
These stellar population models make use of the Padova (1994) 
stellar evolutionary tracks, which do not include the full contribution of the 
Thermally Pulsing - Asymptotic Giant Branch phase, 
as pointed out by Maraston (1998, 2005 hereafter M05) 
and more recently by other authors (e.g. Bruzual 2007, Conroy, Gunn \& White 2008, 
Eminian et al. 2008, Marigo \& Girardi 2007).
As shown in Maraston (1998, 2005), the TP-AGB can account for up to  
$40 \%$ of the bolometric light and $80 \%$ of the light in the K band.
Hence, the neglection of this phase in the galaxy formation models leads to an 
underestimation of the flux 
redwards of $5000 \ \mathring{A}$, with a major effect in the near-IR. 

The emission in this phase is significant for the
stars in the mass range $2-3 \ M_{\odot}$, and therefore peaks in a very
short period of the life of the SSP, around the age of $\sim 1$ Gyr.
For this reason, the TP-AGB light is a powerful marker of an intermediate-age 
stellar population and dominates the K-band luminosity in the post-starburst phases of star formation.
Since the rest of the emission in the K band is contributed 
by old and/or metal rich low-mass stars, the inclusion
of the TP-AGB stars on top of the K-band background leads to dramatic differences
in the colours and the mass-luminosity relation of model galaxies, especially    
in the early phases of their evolution. 

To quantify the effect of the TP-AGB emission on the predicted galaxy
colours at different redshifts, we run a semi-analytical code with
different input SSPs, the M05 models (with TP-AGB) and the PEGASE
models (without TP-AGB), and study the evolution of the
colour-magnitude relation.  For the scope of this paper, the PEGASE
and the Bruzual \& Charlot model SSPs are equivalent, since they are
based on the same input physics. In particular, they make use of the
same stellar evolutionary tracks (Padova tracks) and they employ the
isochrone synthesis modeling, producing very similar SSP spectra (see
M05 for an extended discussion on the properties of these models).  We
use GalICS as our reference semi-analytical model, set with standard
LCDM cosmology and default astrophysical parameters. The model 
includes chemical enrichment and metallicity evolution based
on the instantaneous recycling approximation (for details on the GalICS recipe see Hatton et
al. 2003). We adopt no reddening by dust,
to better isolate the TP-AGB effect.  Between the runs, we do not vary
the input physics except for the different SSP models. The results
presented here do not depend, in essence, on the particular galaxy
formation model, and can be applied to any other model based on CDM
hierarchical clustering (see the Discussion Section).

\section{The predicted colour-magnitude relation}

The evolution of a galaxy population manifests itself in the 
colour-magnitude relation as a function of redshift.
The ageing of the stellar populations, the trend of the SFR with time, 
the metal enrichment, the mass accretion history and the morphology
evolution produce changes in the galaxy spectra and hence in luminosities and colours. 
In our two test runs, the different SSP models are expected to produce
different colour distributions, around the epoch when the 
contribution of the TP-AGB phase to the total emission is maximal. This
is the case when either all the stellar populations are young, or anytime 
the SFR is high enough to provide new stellar populations which, by crossing
the critical age range, overshadow
the background MS and red giant branch (RGB) stars in the near IR with TP-AGB stars.  

\begin{figure*}
\includegraphics[scale=0.95]{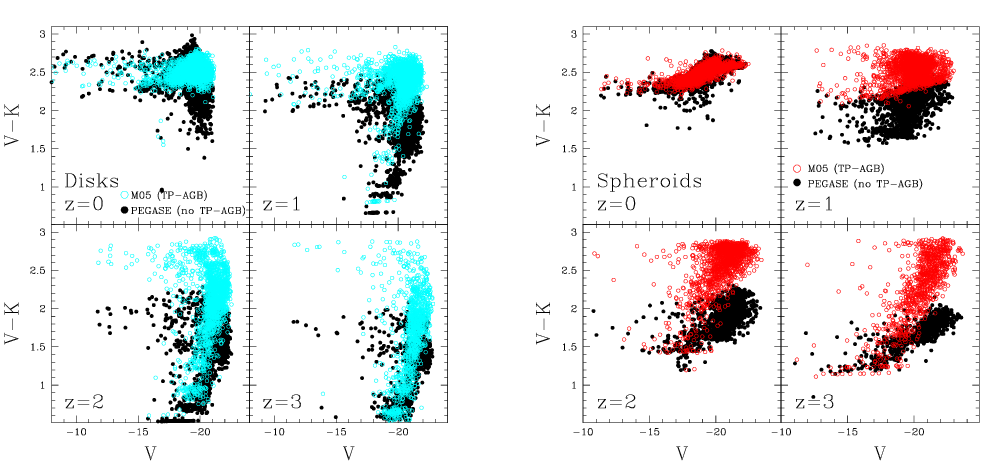}     
\caption{The color-magnitude relation V-K $vs$ V, at redshifts from $z=0$ to $z=3$, 
for disks (\textit{left}) and spheroids (\textit{right}) in a sub-sample 
of the GalICS simulation box (for clarity, a fraction (1/20) of the total ouptut is plotted); 
black filled dots show the PEGASE run, while blue/red empty dots show the 
M05 run. The sample is not complete in magnitude, due to the finite mass resolution of
the GalICS N-body simulation (see Hatton et al. 2003 for details). Here the stellar mass range is 
$2.3e5 M_{\odot} < M_{disk} < 4.5e11 M_{\odot}$ and 
$4.6 e5 M_{\odot} < M_{sph} < 4.5 e11 M_{\odot}$ respectively.}
\label{vkMv}
\end{figure*}

Figure (\ref{vkMv}) shows the rest-frame V-K colour $vs$ the optical V magnitude, 
for 4 different redshifts between $z=0$ and $z=3$, for the two test runs.
The blue/red empty dots represent the M05 run, the black filled dots the PEGASE;
the \textit{left-hand panel} shows the disks population, 
the \textit{right-hand panel} selects the spheroids population, which includes 
ellipticals and the bulges of spirals. For the present scope, the distinctive
feature of these two types of objects is the star formation. 
In disks, the star formation follows the cooling of gas and the SFR is
a function of the gas mass accretion rate and density. 
Spheroids do not directly receive cool gas and form via secular evolution from disks 
or via mergers, and in both cases star formation occurs in bursts. 

Figure (\ref{vkMv}) shows an impressive offset between the two runs.
At $z=3$ the TP-AGB emission in the M05 run produces a reddening of V-K of up 
to $2$ magnitudes, and introduces a spread between the 
blue and the red end of the colour distribution much wider than the PEGASE. 
The distinctive feature of the effect is its dependence with redshift, caused by the
evolution of the star formation rate in the model. 
While at $z=1$ the offset in the colours is still significant, for lower redshifts
it tends to disappear, mirroring the ageing of the stellar populations and the 
fading of the TP-AGB emission into the background IR produced by the MS and RGB 
stars (see M05). 

\begin{figure*}
\includegraphics[scale=0.95]{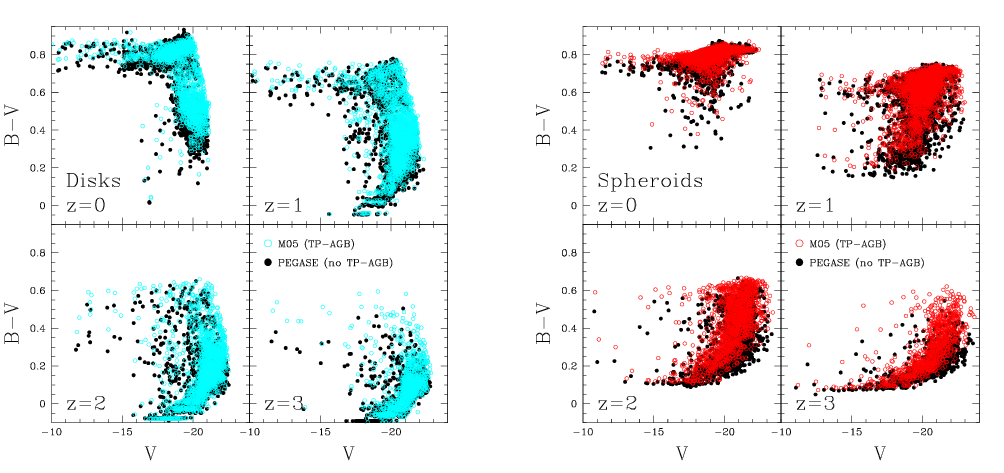}     
\caption{The color-magnitude relation B-V $vs$ V, colour coding as in Fig.~(\ref{vkMv}.)}
\label{bvMv}
\end{figure*}

An even larger offset between the M05 and PEGASE runs  
is shown in the spheroids case (\textit{right-hand panel}). At $z=3$ the two colour
distributions meet only at the faint blue end, while the M05 steeper slope
moves the rest of the objects redwards. 
The offset here is more dramatic, and shows a trend with mass, with the 
most massive galaxies being more affected.
This mirrors the fact that the star formation is more intense
in massive objects, which undergo a higher number of mergers accompanied by
bursts. These events leave no gas available for residual star formation, 
and the massive new stellar populations soon turn into the AGB phase. 
The spheroids are soon dominated by the post-starburst phase, thus
maximising the impact of the red colours. 
In comparison, disks are bluer
than spheroids for a given V-band luminosity and redshift, because of the 
continuous star formation triggered by gas infall, that somewhat dilutes the 
TP-AGB emission. In fact, very young stellar populations are barely affected
by the TP-AGB emission, as already pointed out in Swinbank et al. (2008).    

As expected, the offset decreases as the stellar populations grow old. 
The evolution of the PEGASE colours moves the galaxies from the blue to 
the red end at $z=0$, but at high redshifts the PEGASE run predicts no red objects. 
On the contrary, the M05 run shows a continuous occupation of the red end, which
at high redshifts is contributed by the young populations rich in TP-AGB stars, and 
at low redshift converges to the PEGASE due to the old age of the galaxies. 

It is interesting to compare the M05 and the PEGASE runs in the $B-V$ colours, 
as is shown in Fig.~(\ref{bvMv}). These spectral bands are insensitive to the 
TP-AGB emission, and in fact the two colours distributions are now very similar.
This highlights the main difference between the two runs.  
It is evident that, at any given redshift, an object in the M05 run can be very blue
in $B-V$ and very red in $V-K$, while in the PEGASE run the $V-K$ colours show the
same trend as the $B-V$. This is because, in the M05 run, $V-K$ red colours are not 
associated just with old quiescent galaxies, but are produced also by young objects if 
the TP-AGB phase is active. Notice for instance the blue cloud in $B-V$ at $z=0-1$ for
disks. In the correspondent $V-K$ plot, these galaxies are very red in the M05 run
because of TP-AGB stars, and the resulting colours are virtually undistinguishable
from those of spheroids on the red sequence. This is a remarkable result that 
will be subject to future investigation. 
 
\begin{figure*}
\includegraphics[scale=1.0]{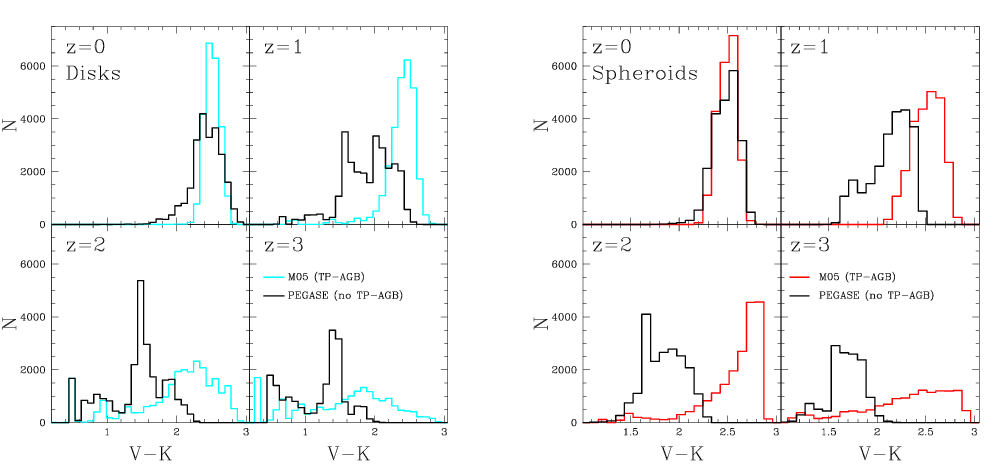}    
\caption{Histogram of the $V-K$ colour distribution for disks and spheroids,
for half the GalICS population ($\sim 23.000$ galaxies at $z=0$), 
colour-coded as in Fig.~{\ref{vkMv}}.}
\label{vkhistogram}
\end{figure*}

Fig.~(\ref{vkhistogram}) portrays the distribution of the $V-K$ colour of 
disks and spheroids for about half the total galaxy population in the 
simulation box ($\sim 23.000$ galaxies at $z=0$), for the M05 (blue/red) 
and PEGASE (black) runs.
In the case of spheroids (\textit{right-hand panel}), 
the two distributions are completely segregated
at high redshift, with an offset of $2$ magnitudes in $V-K$. At $z=1$ the 
offset is still of the order of unity, and disappears at later times. 
For the disks (\textit{left-hand panel}),
the distributions feature a larger spread and an offset of the order of
unity down to $z=1$, and converge at later times.
Fig.~(\ref{vkhistogram}) highlights the different colour evolution of the 
galaxies in the two runs. With respect to the M05, the PEGASE run lacks 
red young galaxies, and is overall bluer at every redshift. The TP-AGB emission
makes the high-redshift M05 galaxies very red, and its effects are still
dominant in $V-K$ down to $z=1$. At the same time, the TP-AGB tends to 
counter-balance a bit of blue light, so that the M05 distribution is
overall more sharply peaked.

\section{The luminosity-mass relation}

\begin{figure*}
\includegraphics[scale=1.0]{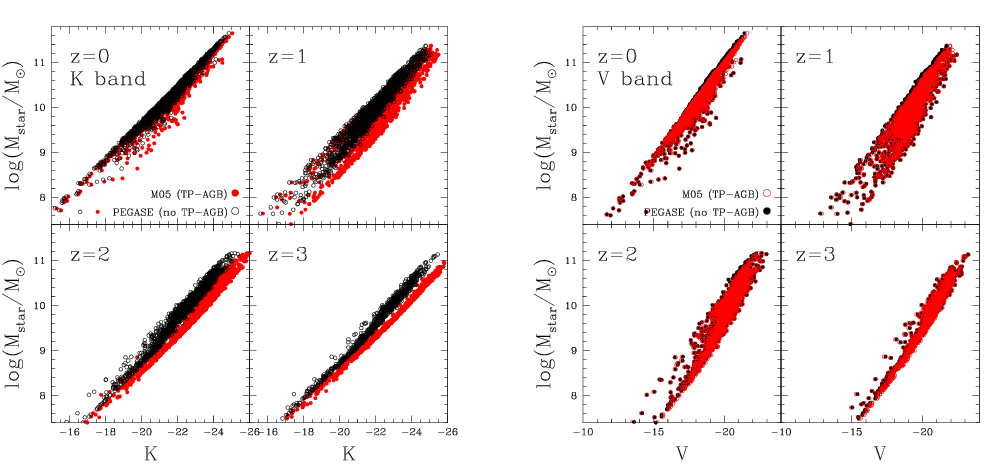}    
\caption{The K-band (\textit{left panel}) and V-band (\textit{right panel}) 
stellar mass $vs$ luminosity relation for the spheroids population, at redshifts 
from $z=0$ to $z=3$ (remnants are taken into account in the stellar mass budget);
colour-coding as in Fig.~{\ref{vkMv}}.}
\label{Mkmbulge}
\end{figure*}

When the TP-AGB emission is not taken into account, the 
total stellar emission in the K band is contributed by low-mass
MS stars and RGB stars with a life span of more than $10$ Gyr.
The correct disentanglement of the TP-AGB from this component 
severely affects the estimate of the mass-to-light ratio ($M/L$), and is
crucial when the K-band luminosity is used as a tracer of the galaxy mass.
In Figure (\ref{Mkmbulge}, \textit{left-hand panel}) 
we show the spheroid mass as a function
of the K-band magnitude, for the same redshifts as the previous plots, and the
same color-coding. 
It is immediately apparent that the offset in the colour-magnitude relation
turns into an offset in the $M/L$ ratio, non-linearly evolving
with redshift. At $z=3$ the mass-luminosity relation for the M05 and PEGASE runs
is again completely separated, touching at the faint blue end and featuring
a different slope. The bias in the K-band luminosity for a given galaxy mass
is more than $1$ magnitude from $z=3$ to $z=1$, with a scatter 
that increases with decreasing redshift, and finally converges at later times. 
For comparison, (\ref{Mkmbulge}, \textit{right-hand panel}) shows the spheroids mass 
as a function of the V-band magnitude, and it is apparent that there is no bias
between the M05 and PEGASE run in this band. This shows that the offset in the 
$M/L$ ratio visible in the K-band is entirely due to the TP-AGB phase.  
 
The predicted K-band magnitude of an elliptical galaxy of $M \sim 10^{11} \ M_{\odot}$ is
$\sim -26$ in the M05 run and $\sim -25$ in the PEGASE.   
A difference of more than $1$ magnitude in K-band luminosity at redshift $3$ is
very significant, especially in the hierarchical structure formation picture. 
To produce a given K-band luminosity without the TP-AGB emission, a higher 
stellar mass is needed, in older stellar populations, and therefore a 
higher star formation rate at high $z$ is required in simulations.   
The amount of time and the number of mergers needed by the 
hierarchical mechanism to produce old massive galaxies at $z=3$ put a 
serious strain on the limits of the theory. The inclusion of the TP-AGB phase, 
by shifting the colours redwards, may help to relieve the hierarchical model of the 
most drastic constraints on the masses of galaxies at high redshifts.

\section{Discussion and conclusions}

The offset in the predicted color-magnitude relation, the galaxy colour distribuition
and their evolution with redshift described in the previous Section
is dramatic, and is entirely produced by different prescriptions 
for the input stellar population models.
The test galaxy formation model GalICS was run with the same input physics in both
cases, ruling out that any feature in the model is responsible for the offset. 
The results presented here are solid against different semi-analytical model or SPH 
simulation implementations, as long as the framework in use is the 
hierarchical growth of structures. The relative 
contribution of the TP-AGB to the overall galaxy emission mainly depends on
the average age of the stellar populations, and therefore is only sensitive
to the mass assembly and star formation history. For this reason we expect similar
results with any galaxy formation model based on CDM hierarchical clustering.
The details of this offset will depend on individual
model recipes, through second-order effects that are expected to
be produced by different implementations of cooling and feedback.

The implications of these results have a great impact on the models 
of structure formation. The redshift-dependent, non-linear offset 
in the predicted K-band luminosities shifts the rest-frame $V-K$ colours redwards 
and the $M/L$ ratio in the red/infrared downwards, especially
for massive galaxies.
The key difference lies in producing extremely red $V-K$ colours in
not-so-massive, young or intermediate-age galaxies, still actively 
star forming and possibly very blue in $B-V$, rather than only to old and dead, 
passively evolving systems. 

It has been pointed out that hierarchical clustering is unable to account for
the observed red galaxy population at high redshifts 
(Cirasuolo et al. 2008, Yan et al. 2004). 
The comparison between observations and simulations 
needs to be reconsidered in the light of the results 
presented here. We produce very red optical to near IR colours  
in young galaxies, without the need of large stellar masses. 
Simulations would be relieved of the constraint to produce
high star formation rates at high redshifts, hard to achieve because 
the dark matter halo assembly process and the baryonic
cooling strain to put together much stellar mass very fast 
at such early epochs. 

We are now set on testing our results 
to see in how far the new model alleviates the difficulties
in the convergence between theoretical predictions and observations.
In follow-up papers we will analyse the impact of our results on key
observational tests such as luminosity function, Tully-Fisher relation
and the properties of the population of extremely red objects. 

\section*{Acknowledgments}
This project is supported by the Marie Curie Excellence Team Grant MEXT-CT-2006-042754
of the Training and Mobility of Researchers programme financed by the European Community.

\end{document}